\documentclass{article}
\usepackage{graphicx}

\newcommand{\ket}[1]{|#1\rangle}
\newcommand{\bra}[1]{\langle #1 |}
\newcommand{\mymid}{\,:\,}
\newcommand{\startproof}{\noindent\emph{Proof.} }
\providecommand{\qed}{\hfill\rule{1.5ex}{1.5ex}}
\newtheorem{lemma}{Lemma}

\newtheorem{proposition}[lemma]{Proposition}

\begin{document}
\title{Conversion of a general quantum stabilizer code to
an entanglement distillation protocol\thanks{%
One page abstract of this paper will appear in
the Proceedings of 2003 IEEE International Symposium
on Information Theory.}}
\author{Ryutaroh Matsumoto\\
Dept.\ of Communications and Integrated Systems\\
Tokyo Institute of Technology, 152-8552 Japan\\
Email: \texttt{ryutaroh@rmatsumoto.org}}
\date{April 4, 2003}
\maketitle
\begin{abstract}
We show how to convert a quantum stabilizer code
to a one-way or two-way entanglement distillation protocol.
The proposed conversion
method is a generalization of those of Shor-Preskill and
Nielsen-Chuang.
The recurrence protocol and the quantum privacy amplification protocol are
equivalent to the protocols converted
from $[[2,1]]$ stabilizer codes.
We also give an example of a two-way protocol converted from a
stabilizer better than the recurrence protocol and the quantum
privacy amplification protocol.
The distillable entanglement by the class of one-way protocols converted from
stabilizer codes for a certain class of states
is equal to or greater than the achievable rate of stabilizer codes
over the channel corresponding to the distilled state,
and they can distill asymptotically more entanglement from
a very noisy Werner state than the hashing protocol.
\end{abstract}
\section{Introduction}
In many applications of quantum mechanics to communication,
the sender and the receiver have to share a maximally entangled
quantum state of two particles.
When there is a noiseless quantum communication channel,
the sender can send one of two particles in a maximally entangled
state to the receiver and sharing of it
is easily accomplished.
However, the quantum communication channel is usually noisy,
that is, the quantum state of the received particle
changes probabilistically from the original state of a particle.

Entanglement distillation protocols \cite{bennett96a}
and quantum error-correcting codes \cite{shor95,steane96b}
are the schemes for sharing a maximally entangled state over
a noisy communication channel.
A distillation protocol is said to be \emph{two-way}
(resp.\ \emph{one-way}) if it involves two-way (resp.\ one-way)
classical communication.
Two-way protocols have larger distillation ability than
one-way protocols.
However, few two-way protocols has been proposed so far,
namely the recurrence protocol \cite{bennett96a}
and the quantum privacy amplification protocol
(QPA protocol) \cite{deutsch96}.
There may be many two-way protocols better than existing ones,
and the discovery of better protocols has been awaited.

Immediately after the proposal
of those schemes,
Bennett et~al.\ discovered that one
can construct a \emph{one-way} entanglement
distillation protocol from a quantum code
\cite[Section V.C]{bennett96}, which requires
$2n$ additional qubits where $n$ is the number of noisy 
entangled states
to be distilled.
Nielsen and Chuang
\cite[Exercise 12.34]{chuangnielsen}
observed a construction method of a one-way protocol without extra qubits
from a real binary quantum stabilizer code
as a generalization of the idea
in \cite{shor00}.

By a conversion method from a quantum code to
a distillation protocol,
we can solve problems of distillation protocols
from results in quantum codes.
For example, we can construct a good distillation protocol
from a good quantum code.
Thus such a conversion method deserves further investigation.

It is not known how one can convert a quantum error-correcting
code to a \emph{two-way} entanglement distillation protocol.
We shall propose a conversion method from
an arbitrary quantum stabilizer code to both one-way
and two-way entanglement distillation protocols
as a generalization of Shor, Preskill \cite{shor00}, Nielsen,
and Chuang \cite{chuangnielsen}.
Benefits of the proposed conversion methods
are
\begin{itemize}
\item We can construct infinitely many two-way protocols.
One can easily construct a two-way protocol
better than the recurrence protocol and the QPA protocol
from a simple stabilizer code (see Section \ref{sec42}).
\item It is known that one-way protocols and quantum error-correcting
codes without classical communication have the same ability
for sharing maximally entangle states over a noisy quantum
channel \cite{bennett96}.
The proposed protocols might be used for further clarification
of the relation between distillation protocols
and quantum error-correcting codes.
\end{itemize}

This paper is organized as follows:
In Section \ref{sec2},
basic notation is introduced.
In Section \ref{sec3},
we present a construction of entanglement distillation protocols
from quantum stabilizer codes.
In Section \ref{sec4},
we give examples of converted protocols equivalent to
the recurrence protocol and the QPA protocol,
and an example better than them.
In Section \ref{sec5}, we evaluate the distillable entanglement by the
class of one-way protocols converted from stabilizer codes,
and show that the converted protocols can distill asymptotically
more entanglement from a noisy Werner state than the hashing protocol
\cite{bennett96}.
In Section \ref{generalfidelity},
we derive a lower bound on fidelity with a general initial state
of protocols.

\section{Notation}\label{sec2}
In this section we fix notation and
the problem formulation.
Let $H_A$ and $H_B$ be $p$-dimensional complex linear spaces
with orthonormal bases $\{\ket{0_A}$, \ldots, $\ket{(p-1)_A}\}$
and $\{\ket{0_B}$, \ldots, $\ket{(p-1)_B}\}$, respectively,
where $p$ is a prime number.
We shall restrict ourselves to $p$-ary stabilizer codes
because an $m$-ary stabilizer code can be constructed as
a tensor product of $p_i$-ary stabilizer codes \cite[p.1831, Remarks]{rains97},
where $p_i$ are prime divisors of $m$,
and extension of the proposed conversion method to the
$m$-ary case is straightforward.
We define the maximally entangled states in $H_A \otimes H_B$ by
\[
\ket{\beta(a,b)} = I\otimes X^a Z^b
\frac{1}{\sqrt{p}}\sum_{i=0}^{p-1}\ket{i_A i_B}
\]
where $a$, $b \in \{0$, \ldots, $p-1\}$,
and matrices $X$ and $Z$ are defined by
\[
X\ket{i} = \ket{i+1 \bmod p}, \;
Z\ket{i} = \omega^i \ket{i}
\]
with a complex primitive $p$-th root $\omega$ of $1$.
The matrices $X$, $Z$ and their commutation relation
were first applied to the quantum mechanics by Weyl \cite[Section 4.15]{weyl31}.
Suppose that Charlie prepares $n$ pairs of particles
in the state $\ket{\beta(0,0)}$,
sends the particles corresponding to $H_A$ to Alice,
and sends the other particles corresponding to $H_B$ to Bob.
The quantum channels between Alice and Charlie and between
Bob and Charlie are noisy in general,
and Alice and Bob share a mixed state $\rho
\in \mathcal{S}(H_A^{\otimes n} \otimes H_B^{\otimes n})$,
where $\mathcal{S}(H_A^{\otimes n} \otimes H_B^{\otimes n})$
is the set of density operators on $H_A^{\otimes n} \otimes H_B^{\otimes n}$.
The state $\rho$ can be an arbitrary density operator.
The goal of an entanglement distillation protocol is
to extract as many pairs of particles with state close to $\ket{\beta(0,0)}$
as possible from $n$ pairs of particles in the state $\rho$.

\section{Protocol}\label{sec3}
In this section we shall describe how to make an entanglement
distillation protocol from a quantum stabilizer code.
In the protocol we extract a state $\tau \in \mathcal{S}(H_A^{\otimes k}
\otimes H_B^{\otimes k})$ from $\rho \in \mathcal{S}(H_A^{\otimes n}
\otimes H_B^{\otimes n})$.

The proposed protocol will be constructed from
the nonbinary generalization \cite{knill96a,rains97}
of quantum
stabilizer codes \cite{calderbank97,calderbank98,gottesman96}.
We assume that the reader is familiar with the formalism of the
nonbinary stabilizer code. Let us introduce notation of stabilizer codes.
Let $E = \{ \omega^i X^{a_1} Z^{b_1} \otimes \cdots
\otimes X^{a_n} Z^{b_n} \mymid$
$a_1$, $b_1$, \ldots, $a_n$, $b_n$, $i$ are integers $\}$,
and $S$ a commutative subgroup of $E$.
The subgroup $S$ is called a stabilizer.

Let $\mathbf{Z}_p = \{0$, \ldots, $p-1\}$ with addition and multiplication taken
modulo $p$.
For a vector $\vec{a} = (a_1$, $b_1$, \ldots, $a_n$, $b_n) \in \mathbf{Z}_p^{2n}$,
let
\[
\mathsf{XZ}(\vec{a}) = X^{a_1}Z^{b_1} \otimes
\cdots \otimes X^{a_n}Z^{b_n}.
\]
Suppose that $\{\mathsf{XZ}(\vec{g}_1)$,
\ldots, $\mathsf{XZ}(\vec{g}_{n-k})$
(and possibly some power of $\omega I$) $\}$
is a generating set of the group $S$,
where $\vec{g}_1$, \ldots, $\vec{g}_{n-k}$
are linearly independent over $\mathbf{Z}_p$.

Let $H$ be a complex linear space with the orthonormal basis
$\{\ket{0}$, \ldots, $\ket{p-1}\}$,
and hereafter we shall identify $H$ with $H_A$ and $H_B$
by linear maps $\ket{i} \mapsto \ket{i_A}$ and
$\ket{i} \mapsto \ket{i_B}$.
Let $Q$ be a stabilizer code defined by $S$,
that is, a joint eigenspace of $S$ in $H^{\otimes n}$.
There are many joint eigenspaces of $S$ and
we can distinguish an eigenspace by its eigenvalue
of $\mathsf{XZ}(\vec{g}_i)$ for $i=1$, \ldots, $n-k$.
Hereafter we fix a joint eigenspace $Q$ of $S$ and
suppose that $Q$ belongs to the eigenvalue $\lambda_i$
of $\mathsf{XZ}(\vec{g}_i)$ for $i=1$, \ldots, $n-k$.

Suppose that we sent $\ket{\varphi} \in Q$, and received
$\mathsf{XZ}(\vec{e}) \ket{\varphi}$.
We can tell which eigenspace of $S$ contains
the state $\mathsf{XZ}(\vec{e}) \ket{\varphi}$ by
measuring
an observable whose eigenspaces are the same
as those of $\mathsf{XZ}(\vec{g}_i)$.
Then the measurement outcome always indicates
that the measured state $\mathsf{XZ}(\vec{e}) \ket{\varphi}$
belonging to the eigenspace  $\lambda_i \omega^{\langle \vec{g}_i,
\vec{e}\rangle}$, where
$\langle \vec{g}_i,
\vec{e}\rangle$ is the symplectic inner product
defined by
\begin{equation}
\langle \vec{g}_i,
\vec{e}\rangle
= \sum_{i=1}^n b_i c_i - a_i d_i,\label{symplectic}
\end{equation}
for $\vec{g}_i = (a_1$, $b_1$, \ldots, $a_n$, $b_n)$
and $\vec{e} = (c_1$, $d_1$, \ldots, $c_n$, $d_n)$.

We define $\vec{g}_i^{\star} = (a_1$, $-b_1$, \ldots, $a_n$, $-b_n)$.
Since the complex conjugate of $\omega$ is $\omega^{-1}$,
we can see that $\mathsf{XZ}(\vec{g}_i^{\star})$ is a
componentwise complex conjugated matrix of $\mathsf{XZ}(\vec{g}_i)$.
Let $S^\star$ be a subgroup of $E$ generated by
$\{\mathsf{XZ}(\vec{g}_1^\star)$,
\ldots, $\mathsf{XZ}(\vec{g}_{n-k}^\star)\}$.
Easy computation shows that $S^\star$ is again commutative.
So we can consider joint eigenspaces of $S^\star$.
There exists a joint eigenspace $Q^\star$ of $S^\star$
whose eigenvalue of $\mathsf{XZ}(\vec{g}_i^{\star})$ is
$\bar{\lambda}_i$ (the complex conjugate of $\lambda_i$).

With those notation,
our protocol is executed as follows:
\begin{enumerate}
\item\label{step1} Alice measures an observable corresponding
to $\mathsf{XZ}(\vec{g}_{i}^\star)$ for each $i$,
and let $\bar{\lambda}_i \omega^{-a_i}$ be the eigenvalue
of an eigenspace of $S^\star$ containing the state
after measurement.
In what follows we refer to $(a_1$, \ldots, $a_{n-k}) \in \mathbf{Z}_p^{n-k}$
as a \emph{measurement outcome}.
\item\label{step2} Bob measures an observable corresponding
to $\mathsf{XZ}(\vec{g}_{i})$ for each $i$,
and let $\lambda_i \omega^{b_i}$ be the eigenvalue
of an eigenspace of $S$ containing the state
after measurement.
In what follows we also refer to $(b_1$, \ldots, $b_{n-k}) \in \mathbf{Z}_p^{n-k}$
as a \emph{measurement outcome}.
\item Alice sends $(a_1$, \ldots, $a_{n-k})$ to Bob.
\item\label{step4} Bob perform the error correction process 
according to $b_1-a_1$, \ldots, $b_{n-k}-a_{n-k}$
as described below.
\item\label{step5} Alice and Bob apply the inverse of encoding operators
of the quantum stabilizer codes.
\item\label{step6} Alice and Bob discards the last $n-k$ particles.
\item\label{step7} If the difference of the measurement outcomes
$(b_1-a_1$, \ldots, $b_{n-k}-a_{n-k})$ indicates that
the fidelity between the remaining $k$ particles
and $\ket{\beta(0,0)}^{\otimes k}$ is low,
Bob discards all of his particles and he tells Alice the disposal
of particles.
\end{enumerate}

We shall introduce some notation.
For a vector $\vec{u}\in \mathbf{Z}_p^{2n}$
let
\[
\ket{\beta(\vec{u})} = (I\otimes \mathsf{XZ}(\vec{u}))
 \ket{\beta(0,0)}^{\otimes n}.
\]
Let $Q(\vec{x})$ [resp.\ $Q^\star(\vec{x})$]
$\subset H^{\otimes n} \simeq H_A^{\otimes n} \simeq
H_B^{\otimes n}$
be the quantum stabilizer code of $S$ (resp.\ $S^\star$)
belonging to the eigenvalue $\lambda_i \omega^{x_i}$ (resp.\ 
$\bar{\lambda}_i \omega^{-x_i}$) of $\mathsf{XZ}(\vec{g}_i)$
[resp.\ $\mathsf{XZ}(\vec{g}_i^\star)$]
for a vector $\vec{x} =(x_1$, \ldots, $x_{n-k})\in \mathbf{Z}_p^{n-k}$,
and $P(\vec{x})$ [resp.\ $P^\star(\vec{x})$]
be the projection onto $Q(\vec{x})$ [resp.\ $Q^\star(\vec{x})$].

\begin{lemma}
We have
\begin{equation}
\{P^\star(\vec{x})\otimes I\} \ket{\beta(\vec{0})} = 
\{P^\star(\vec{x})\otimes P(\vec{x})\} \ket{\beta(\vec{0})} \label{form}
\end{equation}
for any $\vec{x} \in \mathbf{Z}_p^{n-k}$.
\end{lemma}

\startproof
Let $\{\ket{0}$, \ldots, $\ket{p^n-1}\}$ be an orthonormal basis
of $H^{\otimes n}$ consisting of tensor products of
$\{ \ket{0}$, \ldots, $\ket{p-1}\} \subset H$, and we have
\[
\sqrt{p^n}\ket{\beta(\vec{0})} = \sum_{i=0}^{p^n-1} \ket{i}\otimes \ket{i}.
\]
For $\vec{x} \in \mathbf{Z}_p^{n-k}$,
let $\{ \ket{\vec{x},0}$, \ldots,
$\ket{\vec{x},p^k-1}\}$ be an orthonormal basis of $Q(\vec{x})$.
For a state
\[
\ket{\varphi} = \alpha_0 \ket{0} + \cdots
+\alpha_{p^n-1} \ket{p^n-1} \in H^{\otimes n},
\]
we define
\[
\overline{\ket{\varphi}} = \bar{\alpha}_0 \ket{0} + \cdots
+\bar{\alpha}_{p^n-1} \ket{p^n-1},
\]
where $\bar{\alpha}_i$ is the complex conjugate of $\alpha_i$.
With this notation, 
$\{ \overline{\ket{\vec{x},0}}$, \ldots,
$\overline{\ket{\vec{x},p^{n-k}-1}}\}$
is an orthonormal basis of $Q^\star(\vec{x})$.
The set $\{ \ket{\vec{x},i} \mymid \vec{x} \in \mathbf{Z}_p^{n-k}$,
$i = 0$, \ldots, $p^k-1\}$ is an orthonormal basis of $H^{\otimes n}$
and there exists a unitary matrix on $H^{\otimes n}$ that
transforms the basis $\{\ket{0}$, \ldots, $\ket{p^n-1}\}$
to $\{ \ket{\vec{x},i} \mymid \vec{x} \in \mathbf{Z}_p^{n-k}$,
$i = 0$, \ldots, $p^k-1\}$.
Let $\bar{U}$ be the componentwise complex conjugate of $U$,
that is,  $\bar{U}$ transforms $\{\ket{0}$, \ldots, $\ket{p^n-1}\}$
to $\{ \overline{\ket{\vec{x},i}} \mymid \vec{x} \in \mathbf{Z}_p^{n-k}$,
$i = 0$, \ldots, $p^k-1\}$.
We have $\bar{U}\otimes U \ket{\beta(\vec{0})} = \ket{\beta(\vec{0})}$
\cite{horodecki99}.
Therefore
\[
\sqrt{p^n}\ket{\beta(\vec{0})} = \sum_{\vec{x}\in \mathbf{Z}_p^{n-k}}
\sum_{i=0}^{p^k-1} \overline{\ket{\vec{x},i}} \otimes\ket{\vec{x},i}.
\]
Since
\[
P^\star(\vec{x}) = \sum_{i=0}^{p^k-1} \overline{\ket{\vec{x},i}}\,
\overline{\bra{\vec{x},i}},
\]
we have
\begin{eqnarray}
\sqrt{p^n}\{P^\star(\vec{x})\otimes I\} \ket{\beta(\vec{0})} &= &
\left[\sum_{i=0}^{p^k-1} \overline{\ket{\vec{x},i}}\,
\overline{\bra{\vec{x},i}} \otimes I \right]
\sum_{\vec{x}\in \mathbf{Z}_p^{n-k}}
\sum_{i=0}^{p^k-1} \overline{\ket{\vec{x},i}} \otimes\ket{\vec{x},i} \nonumber\\
&=&
\sum_{i=0}^{p^k-1} \overline{\ket{\vec{x},i}} \otimes\ket{\vec{x},i} \nonumber\\
&=&
\sqrt{p^n}\{P^\star(\vec{x})\otimes P(\vec{x})\} \ket{\beta(\vec{0})}\label{form2}
\end{eqnarray}
\qed

Suppose that  we perform the protocol above to the state
$\ket{\beta(\vec{u})} = \{ I\otimes \mathsf{XZ}(\vec{u})\}
 \ket{\beta(\vec{0})}$.
After we get $\vec{a} = (a_1$, \ldots, $a_{n-k})\in \mathbf{Z}_p^{n-k}$
as a measurement outcome in Step \ref{step1},
the state is
\begin{eqnarray*}
&&\{ P^\star(\vec{a})\otimes I\} \{ I\otimes \mathsf{XZ}(\vec{u})\}
 \ket{\beta(\vec{0})}\\
&=& \{ I\otimes \mathsf{XZ}(\vec{u})\} \{ P^\star(\vec{a})\otimes I\}
 \ket{\beta(\vec{0})}\\
&=& \{ I\otimes \mathsf{XZ}(\vec{u})\} \{ P^\star(\vec{a})\otimes P(\vec{a})\}
 \ket{\beta(\vec{0})} \mbox{ [by Eq.\ (\ref{form})]}.
\end{eqnarray*}
Observe that the vector $\{ I\otimes \mathsf{XZ}(\vec{u})\} \{ P^\star(\vec{a})\otimes P(\vec{a})\}
 \ket{\beta(\vec{0})}$
belongs to $Q^\star(\vec{a})\otimes Q(\vec{b})$, where
\[
\vec{b} = \vec{a} + (\langle \vec{g}_1, \vec{u}\rangle,
\ldots, \langle \vec{g}_{n-k}, \vec{u}\rangle).
\]
Thus the measurement outcome in Step \ref{step2} must be $\vec{b}$.

For the simplicity of presentation,
we assume that the state $\rho 
\in \mathcal{S}(H_A^{\otimes n}
\otimes H_B^{\otimes n})$ can be written as
\begin{equation}
\rho = \sum_{\vec{u}\in\mathbf{Z}_p^{2n}}
\alpha(\vec{u}) \ket{\beta(\vec{u})}\bra{\beta(\vec{u})},\label{restrictedstate}
\end{equation}
where
$\{\alpha(\vec{u}) \mymid \vec{u} \in \mathbf{Z}_p^{2n}\}$ is
a probability distribution.
A general case will be treated in Section \ref{generalfidelity}.

After performing Step \ref{step1} in the proposed protocol
to state (\ref{restrictedstate})
and getting $\vec{a} \in \mathbf{Z}_p^{n-k}$ as a
measurement outcome,
the state is
\[
\sum_{\vec{u}\in\mathbf{Z}_p^{2n}}
\alpha(\vec{u})
\{ I\otimes \mathsf{XZ}(\vec{u})\}
 P(\vec{a},\vec{a}) \rho(\vec{0})P(\vec{a},\vec{a})\{ I\otimes \mathsf{XZ}(\vec{u})^*\}
,
\]
where $P(\vec{a}$, $\vec{a}) = P^\star(\vec{a})\otimes P(\vec{a})$ and
$\rho(\vec{0}) = \ket{\beta(\vec{0})}\bra{\beta(\vec{0})}$.
Suppose that we get $\vec{b}$ as a measurement outcome in Step \ref{step2},
and denote $(b_1-a_1$, \ldots, $b_{n-k}-a_{n-k})$
by $\vec{s}$.
The state $\{ I\otimes \mathsf{XZ}(\vec{u})\}
P(\vec{a}, \vec{a}) \ket{\beta(\vec{0})}$ belongs
to $Q^\star(\vec{a})\otimes Q[\vec{a} + (\langle \vec{g}_1$,
$\vec{u}\rangle$, \ldots, $\langle \vec{g}_{n-k}$,
$\vec{u}\rangle)]$.
Thus
the state after Step \ref{step2} is
\begin{eqnarray*}
&&\sum_{\vec{u}\in\mathbf{Z}_p^{2n}}
\alpha(\vec{u})
P(\vec{a},\vec{b}) \{ I\otimes \mathsf{XZ}(\vec{u})\}
P(\vec{a}, \vec{a}) \rho(\vec{0})
P(\vec{a}, \vec{a}) \{ I\otimes \mathsf{XZ}(\vec{u})^*\}
P(\vec{a},\vec{b}) \\
&=&
\sum_{\vec{u} \in D(\vec{s})}
\alpha(\vec{u})
P(\vec{a}, \vec{b}) \{ I\otimes \mathsf{XZ}(\vec{u})\}
 P(\vec{a}, \vec{a}) \rho(\vec{0})
 P(\vec{a}, \vec{a}) \{ I\otimes \mathsf{XZ}(\vec{u})^*\}
P(\vec{a}, \vec{b})
\\
&=&\sum_{\vec{u} \in D(\vec{s})}
\alpha(\vec{u})
\{ I\otimes \mathsf{XZ}(\vec{u})\}
P(\vec{a},\vec{a}) \rho(\vec{0})P(\vec{a},\vec{a})
\{ I\otimes \mathsf{XZ}(\vec{u})\}
,
\end{eqnarray*}
where
\[
D(\vec{s}) = 
\{ \vec{u}\in \mathbf{Z}_p^{2n} \mymid
\langle\vec{g}_i ,\vec{u}\rangle = b_i - a_i,
\mbox{ for each } i\}.
\]

Let $C$ be the linear subspace of $\mathbf{Z}_p^{2n}$
spanned by $\vec{g}_1$, \ldots, $\vec{g}_{n-k}$,
and $C^{\perp}$ be the orthogonal space of $C$
with respect to the symplectic inner product (\ref{symplectic}).
For vectors $\vec{u}$, $\vec{v}$ such that $\vec{u}-\vec{v} \in C$,
$\mathsf{XZ}(\vec{u})$ and $\mathsf{XZ}(\vec{v})$ has the same effect on
states in $Q(\vec{a})$ for any $\vec{a}$,
and we can identify errors $\mathsf{XZ}(\vec{u})$ and
$\mathsf{XZ}(\vec{v})$ if $\vec{u}-\vec{v} \in C$,
which is equivalent to $\vec{v} \in \vec{u}+C$.
Thus, among errors $\mathsf{XZ}(\vec{u})$
corresponding to $D(\vec{s})$,
the most likely error vector $\vec{u}$ is one having maximum
\[
\sum_{\vec{v} \in \vec{u}+C} \alpha(\vec{v})
\]
in the set $D(\vec{s})$.
Let $\vec{e}$ be the most likely error vector in $D(\vec{s})$.
The set $D(\vec{s})$ is equal to
\[
\vec{e}+C^{\perp} = \{\vec{e}+\vec{u} \mymid
\vec{u} \in C^{\perp}\}.
\]

Bob applies $\mathsf{XZ}(\vec{e})^{-1}$ to his particles.
This is Step \ref{step4}.
After applying $\mathsf{XZ}(\vec{e})^{-1}$ to Bob's particles,
the joint state of particles of Alice and Bob is
\begin{equation}
\sum_{\vec{u} \in \vec{e}+C^{\perp}}
\alpha(\vec{u}) \{I\otimes \mathsf{XZ}(\vec{u}-\vec{e})\}
P(\vec{a},\vec{a})
\rho(\vec{0})P(\vec{a},\vec{a})
\{I\otimes \mathsf{XZ}(\vec{u}-\vec{e})^*\}. \label{step4state1}
\end{equation}
Recall that $\mathsf{XZ}(\vec{u}-\vec{e})$ does not change
a state in $Q(\vec{a})$ if $\vec{u}-\vec{e} \in C$.
Therefore the state (\ref{step4state1}) is equal to
\begin{eqnarray}
\lefteqn{\sum_{\vec{u} \in \vec{e}+C}
\alpha(\vec{u}) 
 P(\vec{a},\vec{a})
\rho(\vec{0})P(\vec{a}, \vec{a})+}\nonumber\\*
&&
\sum_{\vec{u} \in \vec{e}+(C^{\perp}\setminus C)}
\alpha(\vec{u}) 
[I\otimes \mathsf{XZ}(\vec{u}-\vec{e})]P(\vec{a},\vec{a})
\rho(\vec{0})P(\vec{a}, \vec{a})
[I\otimes \mathsf{XZ}(\vec{u}-\vec{e})^*].
\label{step4state2}
\end{eqnarray}

We shall explain how to use an encoding operator in Step \ref{step5}
to extract $\ket{\beta(0,0)}^{\otimes k}$ from the above state.
Let $\ket{\mathrm{a}} \in H^{\otimes n-k}$ be an ancillary state.
Consider an encoding operator $U_\mathrm{e}$ on $H^{\otimes n}$
sending $\ket{i} \otimes \ket{\mathrm{a}}\in H^{\otimes n}$
to $\ket{\vec{a},i}$ for $i=0$, \ldots, $p^k-1$, where
$\{ \ket{\vec{a},0}$, \ldots,
$\ket{\vec{a},p^k-1}\}$ is an orthonormal basis of $Q(\vec{a})$
defined above.
Observe that $\overline{U_\mathrm{e}}$
is an encoding operator for $Q^\star(\vec{a})$ sending
$\ket{i} \otimes \ket{\mathrm{a}}\in H^{\otimes n}$
to $\overline{\ket{\vec{a},i}}$ for $i=0$, \ldots, $p^k-1$.
Applying $\overline{U_\mathrm{e}}^{-1}\otimes U_\mathrm{e}^{-1}$
to state (\ref{step4state1}) yields
\begin{eqnarray}
&&\sum_{\vec{u} \in \vec{e}+C^{\perp}}
\alpha(\vec{u}) 
(\overline{U_\mathrm{e}}^{-1}\otimes U_\mathrm{e}^{-1})[I\otimes \mathsf{XZ}(\vec{u}-\vec{e})]P(\vec{a},\vec{a})
\rho(\vec{0})\nonumber\\*
&&\mbox{ }P(\vec{a}, \vec{a})
[I\otimes \mathsf{XZ}(\vec{u}-\vec{e})^*]
(\overline{U_\mathrm{e}}\otimes U_\mathrm{e}) \nonumber\\
& = &
\sum_{\vec{u} \in \vec{e}+C}
\alpha(\vec{u}) 
(\overline{U_\mathrm{e}}^{-1}\otimes U_\mathrm{e}^{-1}) P(\vec{a},\vec{a})
\rho(\vec{0})
P(\vec{a}, \vec{a})
(\overline{U_\mathrm{e}}\otimes U_\mathrm{e})
\mbox{ [by Eq.\ (\ref{step4state2})]} \nonumber\\*
&&\mbox{}+
\sum_{\vec{u} \in \vec{e}+(C^{\perp}\setminus C)}
\alpha(\vec{u}) 
(\overline{U_\mathrm{e}}^{-1}\otimes U_\mathrm{e}^{-1})[I\otimes \mathsf{XZ}(\vec{u}-\vec{e})]P(\vec{a},\vec{a})
\rho(\vec{0})\nonumber\\*
&&\mbox{ }P(\vec{a}, \vec{a})
[I\otimes \mathsf{XZ}(\vec{u}-\vec{e})^*]
(\overline{U_\mathrm{e}}\otimes U_\mathrm{e})\nonumber\\
& = &
\sum_{\vec{u} \in \vec{e}+C}
\alpha(\vec{u}) 
(\overline{U_\mathrm{e}}^{-1}\otimes U_\mathrm{e}^{-1}) 
\left[\frac{1}{p^n}
\left\{ \sum_{i=0}^{p^k-1}
\overline{\ket{\vec{a},i}}\otimes \ket{\vec{a},i}\right\}
\left\{
\sum_{i=0}^{p^k-1}
\overline{\bra{\vec{a},i}}\otimes \bra{\vec{a},i}
\right\}\right]\nonumber\\*
&&\mbox{ }
(\overline{U_\mathrm{e}}\otimes U_\mathrm{e})
\mbox{ [by Eq.\ (\ref{form2})]} \nonumber\\*
&&+
\sum_{\vec{u} \in \vec{e}+(C^{\perp}\setminus C)}
\alpha(\vec{u}) 
(\overline{U_\mathrm{e}}^{-1}\otimes U_\mathrm{e}^{-1})[I\otimes \mathsf{XZ}(\vec{u}-\vec{e})]P(\vec{a},\vec{a})
\rho(\vec{0})\nonumber\\*
&&\mbox{ }P(\vec{a}, \vec{a})
[I\otimes \mathsf{XZ}(\vec{u}-\vec{e})^*]
(\overline{U_\mathrm{e}}\otimes U_\mathrm{e})\nonumber\\
& = &
\frac{1}{p^n}\sum_{\vec{u} \in \vec{e}+C}
\alpha(\vec{u}) 
\left\{ \ket{\beta(0,0)}^{\otimes k}\otimes\ket{\mathrm{a}}^{\otimes 2}\right\}
\left\{ \bra{\beta(0,0)}^{\otimes k}\otimes \bra{\mathrm{a}}^{\otimes 2}
\right\}
  \mbox{ [by definition of $U_\mathrm{e}$]}\nonumber\\*
&&+
\sum_{\vec{u} \in \vec{e}+(C^{\perp}\setminus C)}
\alpha(\vec{u}) 
(\overline{U_\mathrm{e}}^{-1}\otimes U_\mathrm{e}^{-1})[I\otimes \mathsf{XZ}(\vec{u}-\vec{e})]P(\vec{a},\vec{a})
\rho(\vec{0})\nonumber\\*
&&P(\vec{a}, \vec{a})
[I\otimes \mathsf{XZ}(\vec{u}-\vec{e})^*]
(\overline{U_\mathrm{e}}\otimes U_\mathrm{e}) \label{step5state}
\end{eqnarray}
Taking partial trace of the first term over the last $n-k$ qubits
yields $\ket{\beta(0,0)}^{\otimes k}$,
which is Step \ref{step6}.

Let $\tau_5$ be the final state of Step \ref{step5},
that is, state (\ref{step5state}),
and $\tau_6$ be the state after Step \ref{step6}.
In Step \ref{step7},
Bob computes the fidelity between the state
$\ket{\beta(0,0)}^{\otimes k}$ and $\tau_6$
by using knowledge of $\vec{s}$ and $\{\alpha(\vec{u}) \mymid
\vec{u} \in \mathbf{Z}_p^{2n} \}$.
$\mathrm{Tr}[\tau_5]$ is not $1$ because
$\tau_5$ is a state after projection.
We have
\begin{eqnarray*}
\mathrm{Tr}[\tau_6] =\mathrm{Tr}[\tau_5]
&=& \mathrm{Tr}\left[P(\vec{a},\vec{a})
\rho(\vec{0})P(\vec{a}, \vec{a})\right]
\sum_{\vec{u}\in \vec{e}+C^\perp} \alpha(\vec{u})\\
&=&
\bra{\beta(\vec{0})} P^\star(\vec{a})\otimes I
\ket{\beta(\vec{0})} \sum_{\vec{u}\in  \vec{e}+C^\perp} \alpha(\vec{u})
\mbox{ [by Eq.~(\ref{form})]}\\
&=&\frac{1}{p^{n-k}} \sum_{\vec{u}\in  \vec{e}+C^\perp} \alpha(\vec{u})
\mbox{ [by Eq.~(\ref{form2})]}
\end{eqnarray*}

\sloppy
If the initial state is $\ket{\beta(\vec{u})}$ such that
$\vec{u} \in \vec{e}+C$, we can get $(1/p^{n-k})\ket{\beta(0,0)}^{\otimes k}
\bra{\beta(0,0)}^{\otimes k}$
as $\tau_6$. Therefore we have
\[
\bra{\beta(0,0)}^{\otimes k} \tau_6 \ket{\beta(0,0)}^{\otimes k}
\geq 
\frac{1}{p^{n-k}}
\sum_{\vec{u}\in \vec{e}+C} \alpha(\vec{u}).
\]
Thus Bob estimates that the fidelity between $\ket{\beta(0,0)}^{\otimes k}$
and the normalized state of $\tau_6$ is at least
\begin{equation}
\frac{\sum_{\vec{u}\in \vec{e}+C} \alpha(\vec{u})}
{\sum_{\vec{u}\in \vec{e}+C^{\perp}} \alpha(\vec{u})}.
\label{estimatedfidelity}
\end{equation}
The value (\ref{estimatedfidelity}) varies according to
$\vec{s} = (b_1-a_1$, \ldots, $b_{n-k}-a_{n-k})$.
If obtained difference $\vec{s}$ implies low fidelity,
Bob discards all the particles and tell Alice the disposal.
\fussy

Note that if we include Step \ref{step7} then
the whole protocol needs two-way classical communication, but
if we exclude Step \ref{step7} then it needs only
one-way classical communication.

When Alice and Bib do not execute Step~\ref{step7},
the average of fidelity (\ref{estimatedfidelity}) should be
considered instead of respective values of Eq.~(\ref{estimatedfidelity})
for each difference $\vec{s}$ of measurement outcomes.
The average of Eq.~(\ref{estimatedfidelity}) is at least
\begin{equation}
\sum_{\vec{s}\in\mathbf{Z}_p^{n-k}}\sum_{\vec{u}\in \vec{e}(\vec{s})+C}
\alpha(\vec{u})\label{averagefidelity},
\end{equation}
where $\vec{e}(\vec{s})$ is the guessed error vector for a given
difference $\vec{s}$ of measurement outcomes.
This average fidelity (\ref{averagefidelity}) will be studied in
Sections \ref{sec4} and \ref{generalfidelity}.

\section{Examples}\label{sec4}
In this section we show how one can construct the well-known
recurrence protocol and the QPA protocol from stabilizer codes,
and give a two-way protocol constructed from a stabilizer
better than the recurrence protocol and the QPA protocol.

\subsection{The recurrence protocol and the QPA protocol}
The recurrence protocol
without twirling \cite[Step (A2)]{bennett96a}
has the same effect on any density operator
on $H_A^{\otimes 2} \otimes H_B^{\otimes 2}$
as the proposed protocol
with $p=2$, $n=2$, $k=1$, the stabilizer $S$
generated by $Z\otimes Z$, 
encoding operators $U_\mathrm{e}(+1):$
$(\alpha_0\ket{0}+\alpha_1\ket{1})\ket{\mathrm{a}} \mapsto
\alpha_0 \ket{00} + \alpha_1\ket{11}$
for the code belonging to eigenvalue $+1$ of $Z\otimes Z$,
$U_\mathrm{e}(-1):$
$(\alpha_0\ket{0}+\alpha_1\ket{1})\ket{\mathrm{a}} \mapsto
\alpha_0 \ket{01} + \alpha_1\ket{10}$
for the code belonging to eigenvalue $-1$ of $Z\otimes Z$,
and
discarding particles in Step \ref{step7} if
$\vec{s} = (1) \in\mathbf{Z}_2^1$.
This can be seen by a tedious but straightforward computation.

The QPA protocol \cite{deutsch96}
has the same effect as the protocol converted from
the stabilizer $S$
generated by $XZ\otimes XZ$,
encoding operators $U_\mathrm{e}(+1):$
$(\alpha_0\ket{0}+\alpha_1\ket{1})\ket{\mathrm{a}} \mapsto
\alpha_0 (\ket{0}-i\ket{1})(\ket{0}+i\ket{1}) + \alpha_1
(\ket{0}+i\ket{1})(\ket{0}-i\ket{1})$
for the code belonging to eigenvalue $+1$ of $XZ\otimes XZ$,
$U_\mathrm{e}(-1):$
$(\alpha_0\ket{0}+\alpha_1\ket{1})\ket{\mathrm{a}} \mapsto
\alpha_0 (\ket{0}-i\ket{1})(\ket{0}-i\ket{1}) + \alpha_1
(\ket{0}+i\ket{1})(\ket{0}+i\ket{1})$
for the code belonging to eigenvalue $-1$ of $XZ\otimes XZ$,
and
discarding particles in Step \ref{step7} if
$\vec{s} = (1) \in\mathbf{Z}_2^1$

\subsection{A better protocol}\label{sec42}
We shell compare the protocol constructed from
the stabilizer generated by
$\{X\otimes X\otimes X\otimes X$,
$Z\otimes Z\otimes Z\otimes Z\}$ ($p = 2$)
with the recurrence protocol and the QPA protocol
in a similar way to \cite[Fig.~8]{bennett96}.
We discard particles in the protocol unless the measurement outcomes
completely agree, i.e., $\vec{s} = (0,0)$.

Encoding operators for the stabilizer codes
belonging to the eigenvalue $(-1)^{s_1}$ of
$X\otimes X\otimes X\otimes X$ and
$(-1)^{s_2}$ of
$Z\otimes Z\otimes Z\otimes Z$ are
described in Table \ref{tab1}.

Suppose that we have many copies of noisy entangled state
\[
F \ket{\beta(0,0)}\bra{\beta(0,0)}
+
\frac{1-F}{3}
(\ket{\beta(0,1)}\bra{\beta(0,1)}+
\ket{\beta(1,0)}\bra{\beta(1,0)}+
\ket{\beta(1,1)}\bra{\beta(1,1)}),
\]
and we want to distill the Bell state $\ket{\beta(0,0)}\bra{\beta(0,0)}$
as many as possible
by using the hashing protocol and a two-way protocol
chosen from
the recurrence protocol \emph{without twirling},
the QPA protocol, and the protocol
constructed from $\{X\otimes X\otimes X\otimes X$,
$Z\otimes Z\otimes Z\otimes Z\}$.
We use the hashing protocol to distill the perfect the Bell state $\ket{\beta(0,0)}\bra{\beta(0,0)}$ after suitable number of iteration of a two-way
protocol
as described in \cite[Section III.B.1]{bennett96}.

\begin{figure}
\includegraphics*[width=\linewidth]{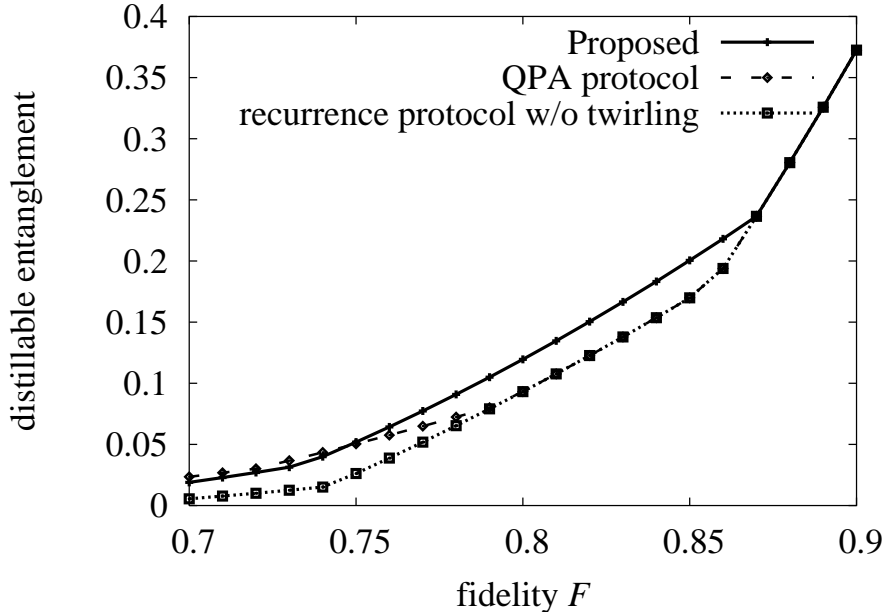}
\caption{Comparison of two-way protocols}\label{fig1}
\end{figure}

The number of perfect Bell state distillable by the three
two-way protocols are compared in Figure \ref{fig1}.
Observe that an example of the proposed protocol has
larger distillable entanglement for the range of $F$
between $0.75$ to $0.87$.

\begin{table}
\caption{Encoding Maps}
\label{tab1}
\[
\begin{array}{|c|c|}
	\hline
	\mbox{eigenvalues}&\mbox{encoding map}\\\hline
	\begin{array}{c}
		(s_1,s_2)=(0,0)
	\end{array}&
	\begin{array}{ccc}
		\ket{00}\ket{\mathrm{a}}&\mapsto&\frac{1}{\sqrt{2}}(\ket{0000}+\ket{1111})\\
		\ket{01}\ket{\mathrm{a}}&\mapsto&\frac{1}{\sqrt{2}}(\ket{0011}+\ket{1100})\\
		\ket{10}\ket{\mathrm{a}}&\mapsto&\frac{1}{\sqrt{2}}(\ket{0101}+\ket{1010})\\
		\ket{11}\ket{\mathrm{a}}&\mapsto&\frac{1}{\sqrt{2}}(\ket{0110}+\ket{1001})
	\end{array}\\\hline
	\begin{array}{c}
		(s_1,s_2)=(0,1)
	\end{array}&
	\begin{array}{ccc}
		\ket{00}\ket{\mathrm{a}}&\mapsto&\frac{1}{\sqrt{2}}(\ket{0001}+\ket{1110})\\
		\ket{01}\ket{\mathrm{a}}&\mapsto&\frac{1}{\sqrt{2}}(\ket{0010}+\ket{1101})\\
		\ket{10}\ket{\mathrm{a}}&\mapsto&\frac{1}{\sqrt{2}}(\ket{0100}+\ket{1011})\\
		\ket{11}\ket{\mathrm{a}}&\mapsto&\frac{1}{\sqrt{2}}(\ket{1000}+\ket{0111})
	\end{array}\\\hline
	\begin{array}{c}
		(s_1,s_2)=(1,0)
	\end{array}&
	\begin{array}{ccc}
		\ket{00}\ket{\mathrm{a}}&\mapsto&\frac{1}{\sqrt{2}}(\ket{0000}-\ket{1111})\\
		\ket{01}\ket{\mathrm{a}}&\mapsto&\frac{1}{\sqrt{2}}(\ket{0011}-\ket{1100})\\
		\ket{10}\ket{\mathrm{a}}&\mapsto&\frac{1}{\sqrt{2}}(\ket{0101}-\ket{1010})\\
		\ket{11}\ket{\mathrm{a}}&\mapsto&\frac{1}{\sqrt{2}}(\ket{0110}-\ket{1001})
	\end{array}\\\hline
	\begin{array}{c}
		(s_1,s_2)=(1,1)
	\end{array}&
	\begin{array}{ccc}
		\ket{00}\ket{\mathrm{a}}&\mapsto&\frac{1}{\sqrt{2}}(\ket{0001}-\ket{1110})\\
		\ket{01}\ket{\mathrm{a}}&\mapsto&\frac{1}{\sqrt{2}}(\ket{0010}-\ket{1101})\\
		\ket{10}\ket{\mathrm{a}}&\mapsto&\frac{1}{\sqrt{2}}(\ket{0100}-\ket{1011})\\
		\ket{11}\ket{\mathrm{a}}&\mapsto&\frac{1}{\sqrt{2}}(\ket{1000}-\ket{0111})
	\end{array}\\\hline
\end{array}
\]
\end{table}

\section{Distillable entanglement by the converted protocols}\label{sec5}
In this section,
we evaluate the distillable entanglement by
one-way protocols constructed from stabilizers.
Distillable entanglement is the most important measure of the performance
of a class of  protocols.

We mean by an $[[n,k]]$ entanglement distillation protocol
a protocol always
leaving $k$ pairs of particles out of given $n$ pairs of particles.
Let $\mathcal{D}$ be a class of $[[n,k]]$ entanglement distillation protocol
for $n=1$, $2$, \ldots, and $k=1$, \ldots, $n$.
Let $\rho_n$ be a density operator on $H^{\otimes 2n}$.
The distillable entanglement by the protocol $\mathcal{D}$
for the sequence of states $\{\rho_n\}$
is the maximum of a real number $R$ such that
for any $R' < R$ and any $\epsilon > 0$ there exists
an $[[n,k]]$ ($k \geq nR'$) protocol in $\mathcal{D}$ such that the protocol
extracts a state $\tau \in H^{\otimes 2k}$ from $\rho_n$
such that the fidelity between $\tau$ and a maximally entangled state
in $H^{\otimes k}$ is at least $1-\epsilon$.
Roughly speaking, the distillable entanglement by $\mathcal{D}$
is the largest number of maximally entangled pairs in $H^{\otimes 2}$ 
distillable from one pair of particles.
Our definition imposes on protocols the restriction that
a protocol always produces the same number of pairs of particles.
A general definition without this restriction was given by
Rains \cite{rains99}.

Let $\{ \alpha(i,j) \mymid (i,j) \in \mathbf{Z}_p^2 \}$
be a probability distribution, and consider the density operator
\[
\rho = \sum_{(i,j) \in \mathbf{Z}_p^2}
\alpha(i,j) \ket{\beta(i,j)}\bra{\beta(i,j)}
\]
on $H_A \otimes H_B$. We shall estimate the distillable entanglement
by the proposed protocol for the sequence of states
$\{ \rho_n=\rho^{\otimes n} \mymid n=1$, \ldots $\}$,
and show the distillable entanglement is at least as large as
the achievable rate of quantum stabilizer codes over
the quantum channel $\Gamma$
on $H$ with an error $X^i Z^j$ occurs
with probability $\alpha(i,j)$.

The achievable rate by quantum stabilizer codes
over $\Gamma$ is the maximum of a real number $R$
 such that
for any $R' < R$ and any $\epsilon > 0$ there exists
an $[[n,k]]$ ($k \geq nR'$) stabilizer code $Q$ such that
any state $\ket{\varphi} \in Q$ can be transmitted over
$\Gamma$ with fidelity at least $1-\epsilon$.

\begin{proposition}
We assume that the decoding of a quantum stabilizer code
is implemented as follows:
First measure an observable whose eigenspaces are the
same as the stabilizer of the code,
determine most likely error of the form
$X^{i_1}Z^{j_1} \otimes \cdots \otimes X^{i_n}Z^{j_n}$,
and apply the inverse of the guessed error to the codeword.
Under this assumption,
the distillable entanglement by the proposed  protocol
without
Step \ref{step7}
for $\{ \rho_n=\rho^{\otimes n} \mymid n=1$, \ldots $\}$
is at least as large as 
the achievable rate by quantum stabilizer codes
over $\Gamma$.
\end{proposition}

\startproof
Let $R$ be the achievable rate by quantum stabilizer codes
over $\Gamma$.
Then for any $R'<R$ and $\epsilon'>0$
there exists an $[[n,k]]$ ($k \geq nR'$)
quantum stabilizer code $Q$ with stabilizer $S$
such that for any state $\ket{\varphi}\in Q$
can be transmitted over $\Gamma$ with fidelity at least $1-\epsilon'$.
Let $S$ be generated by $\{\mathsf{XZ}(\vec{g}_1)$,
\ldots, $\mathsf{XZ}(\vec{g}_{n-k})$
(and possibly some power of $\omega I$) $\}$,
and $Q$ belong to the eigenvalue $\lambda_i$ of $\mathsf{XZ}(\vec{g}_i)$.
Suppose that the decoder guesses the error as $\mathsf{XZ}(\vec{e}(\vec{s}))$
when the measurement outcomes indicate that the received state
belongs to eigenvalue
$\lambda_i \omega^{s_i}$ of $\mathsf{XZ}(\vec{g}_i)$
for $i=1$, \ldots, $n-k$,
where $\vec{s} = (s_1$, \ldots, $s_{n-k})$.
Then the decoder can correct any error $\mathsf{XZ}(\vec{u})$
if
\begin{equation}
\vec{u} \in \{ \vec{e}(\vec{s}) + C \mymid \vec{s} \in \mathbf{Z}_p^{n-k}\},
\label{correctable}
\end{equation}
where $C$ is a linear subspace of $\mathbf{Z}_p^{2n}$
spanned by $\vec{g}_1$, \ldots, $\vec{g}_{n-k}$.

By Lemma \ref{badcodeword} (see Appendix \ref{sec:badcodeword}),
there exists a codeword $\ket{\varphi} \in Q$ such that
if $\ket{\varphi}$ is transmitted and $\mathsf{XZ}(\vec{u})\ket{\varphi}$
is received with $\vec{u}$ not in the set~(\ref{correctable})
then the fidelity between $\ket{\varphi}$ and the decoded state
is at most $9/16$,
because the set~(\ref{correctable}) is equal to the set of correctable
errors by $Q$ in Lemma \ref{badcodeword}.
Since $\ket{\varphi}$ can be transmitted through $\Gamma$ with fidelity
at least $1-\epsilon'$, the probability of the correctable error~(\ref{correctable})
over $\Gamma^{\otimes n}$
is at least $1-16\epsilon'/9$.

Suppose that we apply the proposed protocol to $\rho^{\otimes n}$ such that
if the difference $\vec{s}$ of measurement outcomes is observed
then $\mathsf{XZ}(\vec{e}(\vec{s}))^{-1}$ is applied in Step \ref{step4}.
Then the average (\ref{averagefidelity}) of the fidelity
is at least $1-16\epsilon'/9$,
because the errors in the set (\ref{correctable}) are
also correctable by the proposed protocol [see Eq.\ (\ref{step4state2})].
For given $\epsilon > 0$ set $\epsilon' = 9\epsilon/16$ in the above
argument, and we can see that the distillable entanglement
is at least as large as the achievable rate of quantum stabilizer
codes over $\Gamma$. \qed

The best known lower bound
on the achievable rate by quantum stabilizer codes
over $\Gamma$ is given by Hamada \cite{hamada02}, and his lower bound
gives the true value for the depolarizing channels.
Let us compare the distillable entanglement by the converted protocols
and that by the hashing protocol \cite{bennett96} for
the Werner state of fidelity $F$,
which is given by
$\alpha(1,1) =F$, $\alpha(0,1) = \alpha(1,0) = \alpha(0,0) =
(1-F)/3$ and $p=2$. The Werner state is converted to
\begin{eqnarray}
&&F \ket{\beta(0,0)}\bra{\beta(0,0)} +
\frac{1-F}{3}( \ket{\beta(0,1)}\bra{\beta(0,1)} + \nonumber\\
&& \ket{\beta(1,0)}\bra{\beta(1,0)} +
\ket{\beta(1,1)}\bra{\beta(1,1)}) \label{convertedwerner}
\end{eqnarray}
by applying $XZ$ on Bob's particle.
The distillable entanglement of state (\ref{convertedwerner})
by the hashing protocol
is estimated as
\begin{equation}
1  - H_2(F, (1-F)/3, (1-F)/3, (1-F)/3)\label{binaryhashing}
\end{equation}
where $H_b$ is the Shannon entropy with base $b$.
The distillable entanglement of state (\ref{convertedwerner})
by the converted protocols
is strictly larger than Eq.~(\ref{binaryhashing})
for certain range of $F$,
because the achievable rate of 
the Shor-Smolin concatenated codes is strictly larger than
Eq.~(\ref{binaryhashing}) over the depolarizing channel
of fidelity $F$ \cite{divincenzo98}
and they can be written as stabilizer codes \cite{hamada02}.

Let us consider the case of $p=3$, $\alpha(0,0) = F$,
and $\alpha(i,j) = (1-F)/8$ for $(i,j) \neq (0,0)$.
The distillable entanglement by
the nonbinary generalization \cite{vollbrecht02}
of the hashing protocol is
estimated as
\begin{equation}
1 - H_3(\{\alpha(i,j)\}). \label{ternaryhashing}
\end{equation}
The achievable rate by the quantum stabilizer codes
is strictly greater than Eq.~(\ref{ternaryhashing})
for $0.2552 \leq F \leq 0.2557$ \cite[Section VI.C]{hamada02},
and so is the distillable entanglement by the converted protocols.

\section{Fidelity calculation in general case}\label{generalfidelity}
In the preceding argument we assumed that the initial state
shared by Alice and Bob was in the form of Eq.~(\ref{restrictedstate}).
In this section we remove this restriction.
Let $\rho$ be an arbitrary density operator in
$H_A^{\otimes n}\otimes H_B^{\otimes n}$.
We shall consider applying the proposed protocol without Step
\ref{step7} to $\rho$ and calculate the fidelity
between the distilled state and $\ket{\beta(0,0)}^{\otimes k}$.
Precisely speaking,
we shall calculate the fidelity between
$\ket{\beta(0,0)}^{\otimes k} \otimes \ket{\mathrm{a}}^{\otimes 2}$
and the state after Step \ref{step5},
which is equal to that between $\ket{\beta(0,0)}^{\otimes k}$
and the state after Step \ref{step6}.

The idea of the following argument is borrowed from
Section 7.4 of \cite{preskill98}.
Since there is no selection of particles in
Steps \ref{step1}--\ref{step6} by a measurement,
the whole process of Steps \ref{step1}--\ref{step6}
can be written as a completely positive trace-preserving map
$\Lambda$ on the density operators on
$H_A^{\otimes n}\otimes H_B^{\otimes n}$.

Let $\ket{\psi}\in H_A^{\otimes n}\otimes H_B^{\otimes n}\otimes
H_\mathrm{env}$ is a purification of $\rho$.
Since $\{ \ket{\beta(\vec{x})} \mymid \vec{x}\in \mathbf{Z}_p^{2n}\}$
is an orthonormal basis of $H_A^{\otimes n}\otimes H_B^{\otimes n}$,
we can write $\ket{\psi}$ as
\begin{equation}
\ket{\psi} = \sum_{\vec{x}\in\mathbf{Z}_p^{2n}}
\ket{\beta(\vec{x})} \otimes \ket{\mathrm{env}(\vec{x})},
\label{expansion}
\end{equation}
where $\ket{\mathrm{env}(\vec{x})}$ is a vector in 
$H_\mathrm{env}$.

In Step~\ref{step4}, the inverse error operator
$\mathsf{XZ}(\vec{e})^{-1}$ is determined from
the difference $\vec{s}$ of measurement outcomes and knowledge
of $\{\alpha(\vec{u}) \mymid \vec{u} \in\mathbf{Z}_p^{2n}\}$.
When we deal with an arbitrary but known density operator
$\rho$, determine $\vec{e}$ from $\vec{s}$ so that
the lower bound (\ref{generalbound}) below on fidelity
becomes large.
Once we fix a determination rule of $\vec{e}$ from $\vec{s}$,
we can define $\mathsf{Good} = \{
\vec{u} \in \mathbf{Z}_p^{2n} \mymid$
the protocol can perfectly distill $\ket{\beta(0,0)}^{\otimes k}$
from $\ket{\beta(\vec{u})} \}$.
Equation~(\ref{expansion}) can be written as
\begin{equation}
\sum_{\vec{x}\in\mathsf{Good}}
\ket{\beta(\vec{x})} \otimes \ket{\mathrm{env}(\vec{x})}+
\sum_{\vec{x}\in\mathbf{Z}_p^{2n}\setminus \mathsf{Good}}
\ket{\beta(\vec{x})} \otimes \ket{\mathrm{env}(\vec{x})}.
\label{expansion2}
\end{equation}
The almost same argument as Section 7.4 of \cite{preskill98}
shows that the fidelity between
$\ket{\beta(0,0)}^{\otimes k}$ and the state after
Step~\ref{step6} is at least
\begin{equation}
1 - \left\|
\sum_{\vec{x}\in\mathbf{Z}_p^{2n}\setminus \mathsf{Good}}
\ket{\beta(\vec{x})} \otimes \ket{\mathrm{env}(\vec{x})}
\right\|^2. \label{generalbound}
\end{equation}


\section*{Acknowledgment}
The author would like to thank Prof.\ Tomohiko Uyematsu
and Mr.\ Toshiyuki Morita
for helpful discussions.
This research was supported by
the Japan Society for the Promotion of Science
under contract No.\ 14750278.


\appendix
\section{Bad codeword lemma}\label{sec:badcodeword}
We consider a quantum channel over which
an error of the form $\mathsf{XZ}(\vec{e})$ occurs
with the probability $\alpha(\vec{e})$ for $\vec{e} \in \mathbf{Z}_p^{2n}$,
and we also consider the following decoding method:
Measure the observable of $H^{\otimes n}$
whose eigenspaces are the same as those of $S$,
and apply an operator  $\mathsf{XZ}(\vec{r}_e)$ 
($\vec{r}_e\in \mathbf{Z}_p^{2n}$)
determined by the measurement
outcome and some deterministic criterion.
With this decoding method,
we can correct at most $p^{2n-2k}$ errors
among all the $p^{2n}$ errors for
an $[[n,k]]$ quantum stabilizer code.

\begin{lemma}\label{badcodeword}
Let $Q$ be an $[[n,k]]$ quantum stabilizer code.
Suppose that we have a fixed decoding method
as described above.
There exists a codeword $\ket{\varphi} \in Q$
such that
\[
|\bra{\varphi}\mathsf{XZ}(\vec{r}_e)\mathsf{XZ}(\vec{e})\ket{\varphi}| \leq \frac{3}{4}
\]
for all uncorrectable error $\mathsf{XZ}(\vec{e})$,
where an error $\mathsf{XZ}(\vec{e})$ is said to be
\emph{correctable} if a received state $\mathsf{XZ}(\vec{e})\ket{\varphi}$
is decoded to $\ket{\varphi}$ for all $\ket{\varphi} \in Q$ and
\emph{uncorrectable} otherwise.
\end{lemma}

\startproof
Consider the following map
\[
f:
\left\{
\begin{array}{ccc}
E & \longrightarrow&
\mathbf{Z}_p^{2n}\\
\omega^i X^{a_1}Z^{b_1}\otimes
\cdots \otimes X^{a_n}Z^{b_n} &\longmapsto&
(a_1,b_1,\ldots,a_n,b_n)
\end{array}\right..
\]
Let $C = f(S) \subset \mathbf{Z}_p^{2n}$.
Since $S$ is commutative,
we have $C \subseteq C^\perp$.
Let $C_\mathrm{max}$ be a subspace of
$\mathbf{Z}_p^{2n}$ such that
\begin{eqnarray*}
C_\mathrm{max} &=& C_\mathrm{max}^\perp,\\
C \subseteq & C_\mathrm{max}& \subseteq C^\perp.
\end{eqnarray*}
Such a space $C_\mathrm{max}$ always exists by
the Witt theorem (see Sec.\ 20 of Ref.\ \cite{aschbacher00}).
Since $C_\mathrm{max} = C_\mathrm{max}^\perp$,
we have $\dim C_\mathrm{max} = n$.
The set $f^{-1}(C_\mathrm{max})$ is a commutative subgroup
of $E$, so we can consider a quantum stabilizer code
$Q_\mathrm{min} \subset Q$ defined by $f^{-1}(C_\mathrm{max})$.
We have $\dim Q_\mathrm{min} = p^{n - \dim C_\mathrm{max}} = 1$.
Let $\ket{\psi_1} \in Q_\mathrm{min}$ be a normalized state vector.
We shall construct the desired codeword $\ket{\varphi}$
in Lemma \ref{badcodeword} from $\ket{\psi_1}$.

By the property of stabilizer codes,
if $\vec{x} + C_\mathrm{max} \neq \vec{y} + C_\mathrm{max}$
then
\begin{equation}
\bra{\psi_1} \mathsf{XZ}(\vec{x})^*\; \mathsf{XZ}(\vec{y})\ket{\psi_1} = 0.
\label{orth}
\end{equation}

Let $R \subset C^\perp$ be a set of coset representatives
of $C_\mathrm{max}$ in $C^\perp$, that is,
$R$ has the same number of elements as
$C^\perp/C_\mathrm{max}$, and if $\vec{x}, \vec{y} \in R$
and $\vec{x} \neq \vec{y}$
then $\vec{x} + C_\mathrm{max} \neq \vec{y} + C_\mathrm{max}$.
We assume $\vec{0} \in R$.
Define
\[
\ket{\psi_2} = \frac{1}{\sqrt{p^k}}\sum_{\vec{x}\in R} \mathsf{XZ}(\vec{x}) \ket{\psi_1},
\]
which is a normalized state vector in $Q$
by Eq.~(\ref{orth}).

We want to take $\ket{\varphi}$ in Lemma \ref{badcodeword}
as a multiple of $\ket{\psi_1 + \psi_2}$,
so let us compute
\begin{eqnarray*}
\langle \psi_1 | \psi_2 \rangle &=&
\frac{1}{\sqrt{p^k}}
\sum_{\vec{x}\in R} \langle \psi_1 | \mathsf{XZ}(\vec{x})  | \psi_1 \rangle\\
&=& \frac{1}{\sqrt{p^k}}
\langle \psi_1 | \psi_1 \rangle  \mbox{ by Eq.\ (\ref{orth}) and }
\vec{0}\in R.
\end{eqnarray*}
By Eq.~(\ref{orth}) we also have $\langle \psi_2 | \psi_2 \rangle
= \langle \psi_1 | \psi_1 \rangle$.
Therefore $\langle \psi_1 + \psi_2 | \psi_1 + \psi_2 \rangle = (2 + 
2/\sqrt{p^k}) \langle \psi_1 | \psi_1 \rangle$.
Define $\ket{\varphi}$ by
\[
\frac{1}{\sqrt{2 + 2/\sqrt{p^k}}}\ket{\psi_1 + \psi_2},
\]
which is a normalized state vector in $Q$.
We shall show that $\ket{\varphi}$ has the desired property.

Suppose that an error $\mathsf{XZ}(\vec{e'})$ occurred and
we applied $\mathsf{XZ}(\vec{r}_{e'})$ as the recovery operator.
If $\vec{e} = \vec{e'} - \vec{r}_{e'} \in C$, then
the error $\vec{e'}$ is correctable,
otherwise $\vec{e'}$ is uncorrectable.
If $\vec{e} \notin C^\perp$,
the decoded state is orthogonal to any transmitted state,
so we may assume $\vec{e} \in C^\perp \setminus C$ hereafter.

For $\vec{e} \in C_\mathrm{max}\setminus C$,
\begin{eqnarray*}
&&p^k \bra{\psi_2} \mathsf{XZ}(\vec{e}) \ket{\psi_2}\\
&=&
\sum_{\vec{x},\vec{y}\in R} \bra{\psi_1}\mathsf{XZ}(\vec{x})^*\mathsf{XZ}(\vec{e})
\mathsf{XZ}({\vec{y}}) \ket{\psi_1}\\
&=&
\sum_{\renewcommand{\arraystretch}{0.01}
\begin{array}{c}
\scriptstyle \vec{x},\vec{y}\in R\\
\scriptstyle \vec{x}+C_\mathrm{max} = \vec{e}+\vec{y} + C_\mathrm{max}
\end{array}}
\bra{\psi_1}\mathsf{XZ}(\vec{x})^*\mathsf{XZ}(\vec{e})
\mathsf{XZ}({\vec{y}}) \ket{\psi_1} \mbox{ by Eq.\ (\ref{orth})}\\
&=&
\sum_{\vec{x}\in R} \bra{\psi_1}\mathsf{XZ}(\vec{x})^*\mathsf{XZ}(\vec{e})
\mathsf{XZ}({\vec{x}}) \ket{\psi_1}\\
&=&
\sum_{\vec{x}\in R} \omega^{\langle \vec{e},\vec{x}\rangle}
\bra{\psi_1}\mathsf{XZ}(\vec{x})^*\mathsf{XZ}({\vec{x}}) \mathsf{XZ}(\vec{e})
 \ket{\psi_1}\\
&=&
\bra{\psi_1}\mathsf{XZ}(\vec{e})\ket{\psi_1}\sum_{\vec{x}\in R} \omega^{\langle \vec{e},\vec{x}\rangle}.
\end{eqnarray*}

Consider the linear map $L_{\vec{e}}$
from $C^\perp$ to $\mathbf{Z}_p$ defined by
\[
L_{\vec{e}}(\vec{x}) = \langle \vec{e},\vec{x}\rangle.
\]
Then the kernel of $L_{\vec{e}}$ contains $C_\mathrm{max}$
because $\vec{e} \in C_\mathrm{max}$,
and $\vec{e} \notin C$ implies that $L_{\vec{e}}$ is not a zero
linear map.
Hence we can partition $R$ into cosets of $\ker (L_{\vec{e}})$
in $C^\perp$.
Each coset of $\ker (L_{\vec{e}})$
in $C^\perp$ contains exactly $p^{k-1}$ elements of $R$,
and each element in a coset has the same value under $L_{\vec{e}}$.
Therefore
\begin{eqnarray*}
\sum_{\vec{x}\in R} \omega^{\langle \vec{e},\vec{x}\rangle}
&=&\sum_{\vec{x}\in R} \omega^{L_{\vec{e}}(\vec{x})}\\
&=& p^{k-1} \sum_{i=0}^{p-1} \omega^i\\
&=& 0.
\end{eqnarray*}

Summarizing these results we have
\begin{eqnarray*}
\vec{e} \in C^\perp\setminus C_\mathrm{max} &\Longrightarrow&
\bra{\psi_1} \mathsf{XZ}(\vec{e}) \ket{\psi_1} =0 \mbox{ by Eq.\ (\ref{orth})},\\
\vec{e} \in C_\mathrm{max}\setminus C &\Longrightarrow&
\bra{\psi_2} \mathsf{XZ}(\vec{e}) \ket{\psi_2} =0,
\end{eqnarray*}
and by Eq.~(\ref{orth}) we have for $\vec{e}\in C^\perp$
\[
|\bra{\psi_1}\mathsf{XZ}(\vec{e})\ket{\psi_2}| = \frac{1}{\sqrt{p^k}}.
\]
Thus we have for $\vec{e} \in C^\perp\setminus C$
\begin{eqnarray*}
&&|\langle \psi_1 + \psi_2 |\mathsf{XZ}(\vec{e})| \psi_1 + \psi_2 \rangle|\\
 &\leq&
\frac{1}{2+2/\sqrt{p^k}}
(\underbrace{|\bra{\psi_1}\mathsf{XZ}(\vec{e})\ket{\psi_1}|
+|\bra{\psi_2}\mathsf{XZ}(\vec{e})\ket{\psi_2}|}_{\leq 1} \\
&&\mbox{} +
\underbrace{2 | \bra{\psi_1}\mathsf{XZ}(\vec{e})\ket{\psi_2}|}_{=2/\sqrt{p^k}})\\
&\leq& \frac{1+2/\sqrt{p^k}}{2+2/\sqrt{p^k}}\\
&\leq&  3/4,
\end{eqnarray*}
which completes the proof of Lemma \ref{badcodeword}.
\qed
\end{document}